\documentclass[conference]{IEEEtran}
\IEEEoverridecommandlockouts
\usepackage{cite}
\usepackage{amsmath,amssymb,amsfonts}
\usepackage{algorithmic}
\usepackage{graphicx}
\usepackage{textcomp}
\usepackage{xcolor}
\usepackage{url}
\usepackage{balance}
\usepackage[begintext=``,endtext='',font={itshape,raggedright}]{quoting}
\def\BibTeX{{\rm B\kern-.05em{\sc i\kern-.025em b}\kern-.08em
    T\kern-.1667em\lower.7ex\hbox{E}\kern-.125emX}}
\begin{document}

\title{A grounded theory of Community Package Maintenance Organizations---Registered Report
}

\author{\IEEEauthorblockN{Théo Zimmermann}
\IEEEauthorblockA{\textit{Inria, Université de Paris, CNRS} \\
\textit{IRIF, UMR 8243, F-75013 Paris, France} \\
theo@irif.fr}
\and
\IEEEauthorblockN{Jean-Rémy Falleri}
\IEEEauthorblockA{\textit{Univ. Bordeaux, Bordeaux INP, CNRS} \\
\textit{LaBRI, UMR 5800, F-33400 Talence, France} \\
\textit{Institut Universitaire de France} \\
falleri@labri.fr}

}

\maketitle

\begin{abstract}

\paragraph{Context}


In many programming language ecosystems, developers rely more and more
on external open source dependencies, made available through package
managers.
Key ecosystem packages that go unmaintained create a health risk for
the projects that depend on them and for the ecosystem as a whole.
Therefore, community initiatives can emerge to alleviate the problem
by adopting packages in need of maintenance.

\paragraph{Objective}

The goal of our study is to explore such community initiatives, that
we will designate from now on as Community Package Maintenance
Organizations (CPMOs) and to build a theory of how and why they
emerge, how they function and their impact on the surrounding
ecosystems.

\paragraph{Method}

To achieve this, we plan on using a qualitative methodology called
Grounded Theory.
We have begun applying this methodology, by relying on
``extant'' documents originating from several CPMOs.
We present our preliminary results and the research questions that
have emerged.
We plan to answer these questions by collecting appropriate data
(theoretical sampling), in particular by contacting CPMO participants
and questioning them by e-mails, questionnaires or semi-structured
interviews.

\paragraph{Impact}
Our theory should inform developers willing to launch a CPMO in their
own ecosystem and help current CPMO participants to better understand
the state of the practice and what they could do better.

\end{abstract}


\section{Introduction}

The state of the practice in many programming language ecosystems is
for developers to heavily rely on third-party open source packages\cite{klug_open_2018}.
For instance, Decan \emph{et al.} found that a majority of packages
depend on other packages in all seven ecosystems they
studied~\cite{decan_empirical_2019}.
This is made possible in large part by the advent of package managers,
that have allowed developers to easily add third-party dependencies,
but also to easily share reusable code with others.
This large dependency of many software projects on graciously provided
open source packages can lead to a risk that some of them will be abandoned
and thus stop being maintained~\cite{avelino_abandonment_2019}.
This is especially true for packages with a low truck-factor\cite{avelino_novel_2016,avelino_what_2017}.
Reliance on an unmaintained software package may then create problems
such as the inability to migrate to a newer version of the programming
language or toolchain and reported bugs not being fixed. It can even lead to security issues
as it was the case for the infamous event-stream npm package\footnote{\url{https://blog.logrocket.com/the-latest-npm-breach-or-is-it-a427617a4185/}}.

When an open source package becomes unmaintained, it is possible for
its users to take measures to keep maintaining it, such as pushing
fixes to a fork~\cite{zhou_how_2020}, or vendoring the package in the project that depends
on it, and pushing fixes to this copy~\cite{zimmermann_first_2020}.
However, it is only an individual and uncoordinated measure that will typically 
lead to inefficiencies, as soon as several users need to do the same~\cite{zhou_what_2019}.

To avoid this issue, ecosystem participants may decide to launch
community initiatives to alleviate the problem of key packages being
left unmaintained.
During his PhD, the first author has observed an emerging model of
``community organizations for the long-term maintenance of ecosystems'
packages'' and he has produced an informal analysis mostly based on
the Elm Community
example~\cite{zimmermann_first_2020,zimmermann_challenges_2019}.  His
key observations were that the existence of these organizations could:
facilitate the creation of community forks for unmaintained packages;
provide an exit strategy for authors of popular packages no longer
willing to maintain them.

The goal of the present study is to refine or revise these initial
findings by looking more in depth at several examples of these
Community Package Maintenance Organizations (CPMOs) and build an
actual theory of how and why they emerge, what are their objectives and how they function.

To the best of our knowledge, this theory will be the first formal
study of the CPMO model, which has not been studied by other
researchers so far, and which also constitutes the first known collective model
to alleviate the problem of unmaintained packages in open source
ecosystems.

By providing a formal description abstracting over the many
initiatives that have emerged independently from each other, our
theory will highlight key components and processes of CPMOs, but also
what works well and what does not.
This should allow both practioners from ecosystems without a CPMO to
launch such an initiative, by providing a clear model and
justification for the associated processes, but also current CPMO
participants to reflect on their practice and make them evolve.
As a matter of fact, the first author did launch a CPMO for the Coq
ecosystem based on his initial observations and, even if it has
already been quite successful, it should also benefit from a better
understanding of CPMOs in other ecosystems.

We plan on building this theory through a qualitative study following
the principles of Grounded Theory (GT).
GT is appropriate for this setting because it provides us a
methodology for analyzing both existing data that we retrieve
(``extant documents'') and elicited data (such as through interviews
with CPMO participants).

This study will be of a purely qualitative nature and the theory that
we build will not be used to derive ``predictions'' that could be
``tested''.
It is beyond the scope of this study to make and test such
predictions.
However, the generated theory could inform future research that will
make and test predictions on specific aspects of CPMOs, by means of
quantitative methods.

In the next section, we present the GT methodology and how we have
chosen to apply it to conduct our study.
The rest of the paper then presents our preliminary findings based on
extant documents only, the research questions that have emerged and
how we plan to collect more data to tackle them.

\section{Methodology}

Grounded Theory (GT) is a qualitative methodology for generating
theories grounded in data.
Several variations of this methodology exist. We base our work on the
constructivist version of Charmaz~\cite{charmaz_constructing_2014},
complemented by the perspectives brought by Stol \emph{et al.} on GT
applied to Software Engineering~\cite{stol_grounded_2016} and of
Muller and Kogan on GT applied to Human-Computer Interactions and
Computer-Supported Cooperative Work~\cite{jacko_grounded_2012}. We
also inspire from the SAGE Handbook chapter by Wiener about team work
in GT~\cite{bryant_making_2007} and on the recommendations of Ralph
\emph{et al.} for contextual positioning of extant documents in
GT~\cite{ralph_contextual_2014}.


One of the core characteristics of GT is that it is an incremental
method.
Analysis and theory building (using coding and memoing) start
as soon as the researchers have gathered some initial data and the
resulting theory is then refined by looking at additional data that
will help address unanswered questions.
Data are not sampled for representativity (statistical sampling) but
for what they may bring to the theory under construction (theoretical
sampling).
Data collection only stops when the constructed theory is solid enough
to fit new collected data (theoretical saturation).

\subsection{Defining the scope of our study}

In this paper, we explore a model of community organizations centering
on package maintenance (CPMOs) that we have observed in several
ecosystems.
However, because this model is emergent, we have to define the limits
of what we want to study.
First, we study community initiatives that are rooted in an application-specific
package ecosystem.
Therefore, we exclude both general initiatives targeted at improving
sustainability of open source projects and general-purpose package
ecosystems (such as Arch Linux, Debian, Homebrew or Nixpkgs).
We make this choice because application-specific package ecosystems
are generally associated to an ``upstream'' project, that may become
in need of maintenance, and this leaves the possibility for a CPMO to
``adopt'' the project.
Second, we exclude organizations that do package maintenance without
clearly communicating on their objectives nor their processes.
Third, we exclude organizations that already encompass all community
packages of an ecosystem (as we have sometimes observed in small
ecosystems).
If packages are already gathered like this, this does not leave any
possibility to ``adopt'' an unmaintained package.

\subsection{Initial data gathering}

Our plan for this study has been to start with the data that were the
most accessible, that is the documentation provided by the CPMOs we
identified, and to defer contacting CPMO participants to a later
theoretical sampling phase.

The reason for adopting this strategy is both ethical and practical.
We know that open source software developers are over-sollicited by
empirical software engineering researchers~\cite{baltes_worse_2016}.
Therefore, contacting them too early would be both unethical (because
we would be wasting their time with questions to which we could have
found an answer by ourselves) and inefficient (because reaching out
more personally to specific users is more likely to elicit answers and
we should be able to ask questions which we are missing data to
answer).


Our initial list of CPMOs to study comes from the thesis of the
initial
author~\cite{zimmermann_challenges_2019,zimmermann_first_2020}.
The list was obtained by a systematic search of GitHub organizations
using GitHub advanced search feature, starting from 75 keywords like
``collective'', ``maintain'', ``participate'' but also ``library'',
``module'' or ``package'', and followed by a series of filters
regarding number of repositories, popularity, and presence of
repositories predating the organization (and that were thus
transferred to it).
These filters were intended to keep the resulting list to a size that
would be reasonable to explore manually.
This list was then manually browsed for organizations fitting the
scope described above, excluding in particular many organizations that
did not provide sufficient information on their purpose.

To make sure we did not miss any important or newer CPMO, we completed
this initial list by doing a manual GitHub search for repositories and
organizations with the keywords ``package maintenance'' and ``package
community'' and looking at the first 10 pages of results for each
query.
%
This did bring up additional organizations fitting the scope of our
study, confirming that the list previously established was incomplete.
Even if we still cannot claim to have identified all CPMOs, we have
already obtained a list longer than what we will be able to study in
depth, so we will have to select a subset to focus on.

\begin{table}[h]
\begin{center}
\caption{List of potential CPMOs to study.}
\begin{tabular}{|c|c|c|}
  \hline
  CPMO name & GitHub name & Origin \\
  \hline
  Dlang-community & dlang-community & Initial list \\
  Elm Community & elm-community & Initial list \\
  Flutter Community & fluttercommunity & Manual search \\
  LM Commons & LM-Commons & Manual search \\
  Meteor Community & Meteor-Community-Packages & Manual search \\
  Node.js PMT & pkgjs & Manual search \\
  React-native-community & react-native-community & Initial list \\
  ReasonML-community & reasonml-community & Initial list \\
  Sous Chefs & sous-chefs & Initial list \\
  Vox Pupuli & voxpupuli & Initial list \\
  \hline
\end{tabular}
\end{center}
\end{table}

To start with, we decided to code documents that would present the
CPMO purposes and processes.
We first explored the following four CPMO documents: README from the
manifesto repository of Elm Community, README from the discussions
repository of Dlang-community, README from the package-maintenance
repository of Node.js Package Maintenance Team (PMT), governance
document from the plumbing repository of Vox Pupuli.

We ignored on purpose the coq-community manifesto (which has been
initially drafted by the first author during his PhD work) and the
ocaml-community meta repository README (which has been directly
influenced by coq-community's).
That being said, we do use the experience of the first author as a
comparison point in the analysis process (when writing memos in
particular) and we acknowledge it wherever it influences the theory
under construction.

\subsection{Coding and memoing process}

Following Charmaz's presentation of
GT~\cite{charmaz_constructing_2014}, we have coded these documents in
two phases.
During the initial line-by-line coding phase, we devise codes that
precisely represent the content of the document.
During the follow-up focused coding phase, we abstract our initial
codes and look for common patterns through constant comparison between
codes and codes, codes and data, and data and data.
The focused codes that seem to be the most important become the basis
to form our categories.
Whenever we observe a recurring pattern or gain insight throughout the
coding process, we write memos to sketch the precise definition and
characteristics of our categories.


GT is mainly described as a research methodology employed by
individual researchers (e.g., in sociology).
Guidance on how to employ it as a team of researchers is limited and
it is also the role of the researchers to decide how to proceed.
We decided to use a GitHub repository to collaborate, with issues
being used for writing memos, and documents with codes being committed
into the repository.

Data that are collected but not created for the purpose of the
research are called ``extant
documents''~\cite{charmaz_constructing_2014}.
Extant documents are harder to code because they do not only contain
data that are relevant to our research (these data are typically
burried under irrelevant ecosystem-specific technical considerations
for instance).
To alleviate this difficulty, we decided to code documents separately
and compare our codes.

While such double-coding is generally conducted to obtain more
``objective'' results in empirical software engineering research (by
measuring inter-rater reliability), it is typically not required from
a constructivist perspective (where it is expected that different
researchers will have different interpretations).
However, this was still very useful to us because the differences in
our codes raised interesting discussions during our meetings, and
frequently led to memo-writing.
This was especially true when comparing focused codes, so we quickly
limited (after two documents) the double-coding and discussion to the
focused coding phase only, or we adopted a strategy where one
researcher would do the initial coding and another would do the
focused coding (of the same document) then the two would discuss the
focused codes.
This strategy aligns with the observations of
Wiener~\cite{bryant_making_2007} that team meetings can be used to
code as a group and may spark new ideas leading to memos that will be
written by individual researchers.

When coding extant documents, it is important to situate them with
respect to their context, audience, etc.
Ralph \emph{et al.}  call this ``contextual positioning'' and provide
a list of questions to ask about the
document~\cite{ralph_contextual_2014}.
In order to answer them, in particular with respect to community
documents that are not attributed to a specific author but were
committed to a GitHub repository, we rely on the git history to better
understand who contributed to writing what parts and when.
This is sometimes helpful to explain how different and apparently
contradictory perspectives coexist in the same document, or how CPMO
processes evolve.

\subsection{Theoretical sampling}


After our initial data gathering and analysis, we have already started
theoretical sampling to answer questions that have emerged.
We were especially interested in the package ``adoption'' process, so
we have started coding adoption discussions in the context of one
specific CPMO (Dlang-community) but we plan to expand this to other
CPMOs.

The adoption process was not really discussed in Node.js PMT's
document and while looking for adoption discussions, we figured out
that it was still undecided if this organization (or Working Group as
they sometimes describe themselves) would ever adopt packages.
Currently, they are more focused on developing tooling and best
practices, and helping projects outside the organization.
While this initiative is also very interesting and would be worth
studying, we have to set limits to what we include in our study, so we
have decided to remove it from our study and reinspect our preliminary
findings based on this decision.

Finally, when we reach the limits of what we can theorize from extant
documents, we believe that we will need to contact CPMO participants.
Contacting and interviewing participants is essential, not only to
gain a deeper understanding of the studied processes, but also to get
a reliability check~\cite{stol_grounded_2016} of our (intermediate)
conclusions.

One strategy for contacting CPMO participants will be to e-mail them
and ask them specific questions to clarify the data that we have found
and analyzed.
An example is asking why a package adoption did not happen even after
it was discussed and approved, or if they can tell us under what
circumstances the documented decision processes are deviated from.

When sending these e-mails, we will use the opportunity to ask about
conducting an interview with them, but we will leave it open to have a
purely written discussion for those who would not have time to
allocate to a proper interview.

\subsection{Interviewing process}

During this study, we expect that we will conduct some semi-structured
interviews.
Given the incremental nature of GT, we plan to conduct interviews one
by one, transcribe and code the interview, and reflect on the
knowledge we gain from an interview before planning the next one.
Therefore, the interview guide will evolve between each interview.

Whereas, as an open science committment we will publish all our codes
for public documents, we will not be able to do the same for interview
transcripts and codes.
Indeed, we can expect participants to answer differently to our
questions if we tell them that the interviews will not be made public.
We plan to quote some interview participants in our paper, but we will
seek consent from each quoted participant for the specific quotes, to
make sure that they are comfortable making these quotes public (or if
they want them sanitized and anonymized) and also that we interpret
them correctly.


\subsection{Expectations for the Registered Report}

Registered Reports were initially invented to avoid HARKing
(hypothesizing after the results are known).
This issue is not really relevant for qualitative theory-generation
studies.
Nonetheless, preregistrations can still be useful in an open science
perspective to state the plans and intents for a qualitative study.
Revealing these will help future readers understand how the study was
conducted and the theory was built, even though, in a GT setting, it
is expected that there will be variations from the initial plan.

Furthermore, early reviewer feedback, both on the methodology and the
theory under construction, is essential to us before we actually start
contacting participants and conduct interviews.
Indeed, as we stated before, the time of open source maintainers (and
thus CPMO participants) is precious and there is only so much time
that they can allocate to researchers interviewing or surveying them.
Therefore, it is important that we are as prepared as possible before
we begin this step in our study.


\section{Preliminary results}

In this section, we describe our preliminary categories, obtained after our initial 
and focused coding phases performed on the manifesto documents of the subjects 
we gathered.

\subsection{Motivations for CPMOs}

Avoiding ending-up with lagging useful packages is the main driver for the creation of CPMOs.
Both the Elm Community and the Dlang-community manifestos express this
idea (in words that come from the initial version of the documents by
the CPMO founders):

\begin{quoting}
    It sometimes happens that packages which are widely used need a bit of
    maintenance---for instance, to accommodate changes in Elm, or for
    other reasons. Normally, package authors will deal with that
    themselves, of course, possibly with the help of pull requests from
    interested community members etc. However, sometimes package authors
    may not be available, for one reason or another, and other work can be
    blocked until the maintenance is performed.
\end{quoting}

\begin{quoting}
    This organization was formed by annoyance of needing to fork popular
    repositories to get fixes merged. [...] However, small bug fixes don't
    need to wait in the queue for months, and in case the author is
    completely gone, the DLang community has one upstream repository
    instead of ten different forks containing the same fix.
\end{quoting}

The Vox Pupuli webpage also expresses this idea:

\begin{quoting}
    One of the benefits we hope to achieve is that by a shared ownership
    of modules we no longer end up in situations where the original
    maintainer has moved on and a forest of disparate forks try to fill
    the void.
\end{quoting}

\subsection{Package adoption}

To avoid ending-up with lagging packages, CPMOs perform \emph{package adoptions}, by transferring or
forking the project of interest inside a CPMO-owned repository.
However, we have seldomly encountered clear \emph{guidelines} on how
the packages to adopt are selected.
The more informative description is perhaps the one found in Dlang-community, as follows.

\begin{quoting}
These question are intended to give a rough feeling on what packages might be considered for adoption. When in doubt, please open an issue! \\
Popularity \\
Q: Is there enough interest from the D community, i.e. is it ``worth maintaining'' \\
Competition \\
Q: Is there a similar library with active development out there? \\
Maintainer/sponsor \\
Q: Is at least one of the DLang community member competent for the domain covered by the project? If not, is there anyone willing to join?
\end{quoting}

Here we can notice that the dimensions are popularity, competition and having a volunteer maintainer. Furthermore, there is an explicit will to ensure that adoption is to be discussed prior to any decision.

\begin{quoting}
Please don't create new packages without consulting other dlang-community members.
\end{quoting}

In the Elm Community CPMO, the guidelines are more fuzzy.
It has a mention in the framing of the manifesto document:

\begin{quoting}
packages that are widely used
\end{quoting}

Which indicates that popularity could be a criterion similarly to Dlang-community. Additionally, they state:

\begin{quoting}
If you think there is a package that should be adopted here, feel free to open an issue in this repository. (And, to repeat, this is not to discourage you from adopting a package yourself, if you want to take that on).
\end{quoting}

Which does seem very inclusive as every adoption proposal can be
subjected to discussion.

No criteria for package adoption are explicitly mentioned in the Vox
Pupuli CPMO (they only provide technical adoption guidelines).
However, Vox Pupuli documentation hints that \emph{donations} are a
way to adopt packages, which could bear some similarity with the
vision of Dlang-community (is there a volunteer maintainer out
there?).

\subsection{Maintenance objectives of CPMOs}

The \emph{maintenance objectives} of CPMOs with respect to the adopted
packages are generally not clearly defined.
A theoretical sensitizing concept here is the staged model of the
software lifecycle by Rajlich and Bennett~\cite{rajlich_staged_2000}.
Do CPMO generally plan to do servicing only or do they also accept
projects that are actively evolving or even in their initial
development phase?
The Elm Community manifesto is the most explicit on this aspect, but
it is also self-contradicting as shown by these two excerpts:

\begin{quoting}
For the most part, we don't expect to do innovative work on packages
here. That is, to the extent that innovative new features can be added
to a package, that should mostly be done in people's individual
accounts. What we'll do here is mostly maintenance.
\end{quoting}

\begin{quoting}
Lead the direction of a repository. As a champion, you will need to
make calls on API design. Don't let packages come to elm-community to
die.
\end{quoting}

This apparent contradiction can be explained by the latter paragraph
having been added by someone else than the original author, who
probably had slightly differing views.

\subsection{Decision protocols}

CPMO processes require frequent decision making.
Examples of important decisions to make are whether to ``adopt'' a package or whether to integrate a pull request.

CPMOs often strive to involve the community in the decisions as much as possible.
On the other hand, they also want to be able to make decisions in a reasonable time-frame.
One typical trade-off is the use of a \emph{lazy consensus} protocol,
as in Vox Pupuli.
As written in the Vox Pupuli governance document:

\begin{quoting}
Lazy consensus is a very important concept within the project. It is this process that allows a large group of people to efficiently reach consensus, as someone with no objections to a proposal need not spend time stating their position, and others need not spend time reading such statements.
\end{quoting}

By exploring the file history for this document, we have found out
that this text comes verbatim from a template for open source project
governance provided by OSS Watch~\cite{gardler_meritocratic_2013}.



%


From our preliminary observation, it is not clear at all that the
recommendations of this consensus mechanism are often followed, or for
which type of decision they are applied.

Dlang-community does not explicitly provide a lazy consensus protocol,
but does rather provide some criteria for when it is not required:

\begin{quoting}
When can I (self)-pull a PR?
\end{quoting}


Finally, Elm Community is more explicit on how individual hosted
projects are managed: all projects have a \emph{primary maintainer}
that can take all decisions with respect to its projects.
However, to ensure the liveliness of the maintenance, they state that
unresponsive maintainers can be overriden (and even removed).

\begin{quoting}
if there's something that really needs to get merged, and the maintainer has taken more than 7 days to respond, we can merge things without their involvement.
\end{quoting}

\begin{quoting}
Unresponsive champions will be emailed. Lack of response will mean a
new maintainer will be assigned to that repo.
\end{quoting}

Even if it is not documented as clearly, a question is whether this
principal maintainer model happens to be prevalent in CPMOs in
practice.
This may turn out to be true at least for projects whose original
author is still active, as hinted by Dlang-community:

\begin{quoting}
projects are still driven by their original authors if they have the time
\end{quoting}

\subsection{CPMO membership}

CPMOs generally encourage \emph{wide participation}.

\begin{quoting}
How do I become a member of the DLang community?
First of all, by reading this you most likely are already.
\end{quoting}

\begin{quoting}
Contributors \\
How to become one: Submit a pull request to a Vox Pupuli project
\end{quoting}

However, they also often have a \emph{trust-building process} to get
write permissions:

\begin{quoting}
You are already a well-known member of the D community, then simply ping us for merge rights. \\
Otherwise, start contributing to one of the projects and earn your trust.
\end{quoting}

\begin{quoting}
Collaborators are contributors who have shown wide dedication to the
Vox Pupuli project in general or deep dedication to one project in
particular, and the ability to work well with contributors and other
users.
\end{quoting}

Interestingly, while this last quote also originally comes from the
governance template mentioned above, it was adapted to the specifics
of a CPMO.
Indeed, contrary to usual open source projects, involvement in a CPMO
has several \emph{dimensions}: it can be \emph{wide} dedication to the
CPMO, by helping maintain all packages or common tools, or it can be
\emph{deep} dedication to one or a selection of packages hosted by the
CPMO.

Finally, our preliminary observations seem to indicate that the
trust-building process can often be side-stepped when \emph{recruiting
a new member} to maintain a package being adopted (as shown by these
quotes from Dlang-community, Vox Pupuli and Elm Community):

\begin{quoting}
If not, is there anyone willing to join?
\end{quoting}

\begin{quoting}
It is also common to give collaborator status to an individual who
donates code to the project by migrating a repository to the github
namespace
\end{quoting}

\begin{quoting}
this is not to discourage you from adopting a package yourself, if you
want to take that on
\end{quoting}

\section{Next steps}

In this section, we describe our planned next iterations of data collection and coding, relatively to the categories described previously.

\paragraph*{Motivations of CPMOs}

The manifestos we coded are relatively scarce about the rationales and
motivations that led to the creation of CPMOs.
Therefore, we plan to collect more data about the creation context of
CPMOs.
For several subjects, we gathered some forum threads where the initial
discussion regarding the CPMO creation was conducted.
We plan to code these documents.

When we cannot find data on the creation context or if we need
additional information, we plan to interview CPMO founders.
In the case of Elm Community, we already have the recording of the
informal interview of the founder conducted by the first author during
his PhD.
We will transcribe and code this interview.

\paragraph*{Adoption of packages}

As we previously described, CPMOs are rarely explicit about the
criteria for package adoption.
Even for a CPMO that provides them, these criteria are very
subjective.
We feel like we do not have enough data to construct a comprehensive
theory about this topic.
During our inspection of the CPMO repositories, we found out that
several CPMOs discuss package adoptions using GitHub issues or pull
requests (or mailing lists).
We think that coding these data is a great way to better analyze how
the decision on a package adoption is done in practice.
We already started coding five such issues for the Dlang-community
CPMO, and we plan to conduct this coding effort for the other CPMOs
where such data are available.
When the data are not available, or when we need additional
information, we will ask questions regarding the adoption criteria to
current CPMO participants.

\paragraph*{Maintenance objectives of CPMOs}

We uncover contradictory visions inside CPMO manifestos regarding the
kind of maintenance expected in adopted packages.
We want to examine data closer to the projects to build a better
understanding.
Therefore, we plan to code commit logs of several adopted projects
inside each CPMO.
Because commit messages are usually quite explicit about the kind of
maintenance activity, we think that this should help understand what
kind of maintenance activities are undertaken in CPMOs, and if they
vary across hosted projects.
Contrary to what is commonly seen in research coding commit messages,
we will not do this with quantitative / statistical objectives, but
only to assess the diversity of undertaken maintenance activities.

\paragraph*{Decision protocols}

The studied CPMOs have defined two kinds of decision protocols: lazy
consensus or principal maintainer lock.
However, we have not yet examined how these protocols are used in
practice, and for which reasons.
To build a better understanding, we plan to code issues and pull
requests of adopted projects' repositories as well as issues and pull
requests of central documentation or coordination CPMO repositories.
Additionally, some CPMOs have public discussion channels that are
also used to make decisions.
We plan to also code a subset of these discussions.

\paragraph*{CPMO membership}

Besides an expressed intent to favor wide participation, CPMOs
generally do not precisely document the actual recruitment and trust
building process.
We will try to better understand how this happens in practice based on
public data, but this will very likely need to be complemented by
interviewing both recently recruited participants and CPMO
administrators.
We will also try to analyze the variety of (formal or informal) roles
taken by CPMO participants, and to categorize them along the
\emph{dimensions of involvement} defined above.

\balance
\bibliographystyle{IEEEtran}
\bibliography{biblio}

\end{document}